%% file: sample-sigconf.tex
\documentclass[sigconf]{acmart}
\usepackage{array}
\AtBeginDocument{%
  }

\setcopyright{cc}
\setcctype{by}
\copyrightyear{2026}
\acmYear{2026}
\acmDOI{10.1145/3776574.3831126}
\acmConference[ICMI '26]{INTERNATIONAL CONFERENCE ON MULTIMODAL
  INTERACTION}{October 05--09, 2026}{Napoli, Italy}
\acmISBN{979-8-4007-2318-6/26/10}

\begin{document}
\title{CHORUS: Designing Human–AI Multi-Agent Collaboration for Professional Translators}

\author{George Xi Wang}
\authornote{These authors contributed equally to this work.}
\email{xw3617@nyu.edu}
\affiliation{%
  \institution{New York University}
  \city{Brooklyn}
  \state{New York}
  \country{United States}
}
\affiliation{%
  \institution{Stony Brook University}
  \city{Stony Brook}
  \state{New York}
  \country{United States}
}

\author{Jiaqian Hu}
\authornotemark[1]
\email{jiaqianh@middlebury.edu}
\affiliation{%
  \department{Translation and Localization Management}
  \institution{Middlebury Institute of International Studies at Monterey}
  \city{Monterey}
  \state{California}
  \country{United States}
}

\author{Guande Wu}
\email{guandewu@nyu.edu}
\affiliation{%
  \institution{Tandon School of Engineering, New York University}
  \city{New York City}
  \state{New York}
  \country{United States}
}

\author{Jing Qian}
\email{jqian1590@tongji.edu.cn}
\affiliation{%
  \department{College of Electronic and Information Engineering}
  \institution{Tongji University}
  \city{Shanghai}
  \country{China}
}

\renewcommand{\shortauthors}{Wang et al.}

\begin{abstract}
  \input{Sections/abstract}
\end{abstract}



\begin{teaserfigure}
  \centering
  \includegraphics[width=0.8\textwidth]{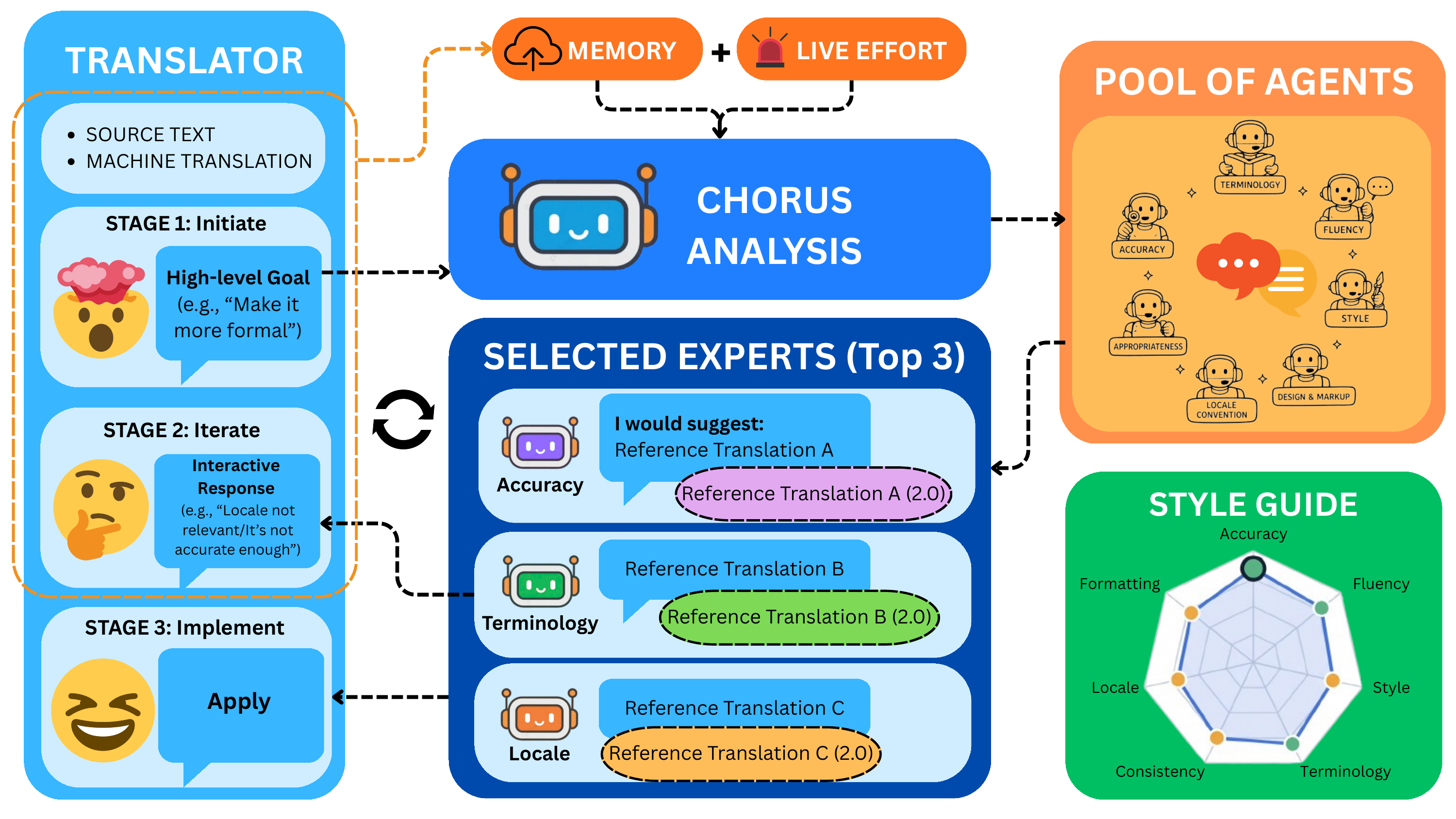}
  \caption{CHORUS is a multi-agent translation system for professional translators. With seven MQM-inspired agents, it provides concurrent editing suggestions while adapting to translators' editing habits from multi-modal inputs and editing histories. }
  \Description{Workflow diagram of the CHORUS multi-agent translation pipeline. Starting from the source text and a machine translation, the translator states a high-level goal, iterates on interactive responses, and applies suggestions, while a central CHORUS analysis module, informed by memory and live effort signals, selects the top three experts from a pool of quality-focused agents, each proposing a candidate reference translation. A style guide panel with a seven-axis radar chart captures the translator's evolving preferences.}
  \label{fig:teaser}
\end{teaserfigure}

\maketitle

\input{Sections/01_introduction.tex}

\input{Sections/02_related.tex}
\input{Sections/03_formative_study}
\input{Sections/04_system}
\input{Sections/05_user_study}
\input{Sections/06_results}
\input{Sections/07_discussion}
\input{Sections/08_limitation_n_future}
\input{Sections/09_conclusion}

\input{Sections/10_acknowledgement}
\bibliographystyle{ACM-Reference-Format}
\bibliography{references}


\end{document}

%% file: Sections/abstract.tex
Despite the widespread use of automatic AI translation systems in daily language tasks, their limitations become apparent in professional translation contexts, where human expertise remains crucial. Professionals rarely rely on these systems in their practice due to a lack of detailed support for the translation process, matching professional styles, and accountability for the final outcome. To bridge this gap, we present CHORUS, a mixed-initiative system designed to support translators' workflow while preserving their personal style. The system adapts through behavioral interaction signals, such as keystrokes, pauses, and editing effort. A formative study found that incorporating MQM theory may be beneficial for professional translation, and the system should adapt to each individual translator's idiosyncratic traits. The final within-subject study with 30 licensed English--Chinese translators found that our system reduced completion time by 33.8\%, lowered translators' cognitive effort, and improved final translation quality measured by the automatic metrics BLEU and COMET. Participants also reported that the system made translation issues easier to inspect, reduced repeated prompting compared to a chat-interface LLM, and offered reflections on their habits and traits. Our findings illustrate how multi-agent AI systems can be designed to support expert workflows and their potential for professional use. 

%% file: Sections/01_introduction.tex
\section{Introduction}
Professional translation is a crucial part of high-stakes scenarios, such as medical, legal, and business communications, where errors are much less tolerated, and quality and style are more prominent than in daily translations. Large Language Models (LLMs) have become increasingly popular, but they often lack accurate meaning preservation, cultural nuance, domain specificity, and personal style~\cite{jiao2023chatgpt,ye2025well,zhang2025good,briva2024large,rivas2025translators,pang2025salute} for professionals. Moreover, professional translation needs to meet specific requirements imposed by clients, and such requirements are rarely available in public domains for LLMs to learn. Translators often need to discuss and resolve nuanced details with clients of different cultural backgrounds, as they are solely responsible for the final outcomes. This responsibility is particularly critical in high-stakes domains where human assurance and accountability are still essential. Plus, most existing systems use a single LLM interface, which often blends accuracy, terminology, and style into a one-shot translation. This makes it difficult for translators to revise any of these dimensions individually, resulting in additional time during post-editing. Consequently, existing systems fail to provide the in-progress granularity needed to support professional translation, and guidance in the literature on how to build such systems is largely missing. As a result, we ask: how can we harness the power of LLMs to design an efficient tool to support professional translation work?

Through a formative evaluation with 6 professional translators, we gather design insights around how to better scaffold professional work by incorporating Multidimensional Quality Metrics (MQM), a framework widely used to describe translation quality across dimensions such as accuracy, terminology, fluency, and style~\cite{lommel2014multidimensional,lommel2018metrics,freitag2021experts}. Rather than treating MQM as an evaluation taxonomy, we use it to structure support during revision and enable translators to inspect, adjust, and review each dimension.

We present CHORUS, a mixed-initiative translation system that adopts an MQM-inspired multi-agent design. We develop a live effort algorithm that uses multi-modal input and editing history to adjust agents' focus during the working process. A Live Style Guide summarizes revision patterns and provides personalized feedback for idiosyncratic traits. This design goal raises three research questions: \textbf{RQ1:} whether CHORUS reduces translators' editing effort and perceived workload; \textbf{RQ2:} whether it improves final translation quality; and \textbf{RQ3:} how CHORUS supports professional translators in revision, reflection, and preference formation.

We evaluated CHORUS in a within-subject study with 30 licensed English--Chinese translators using WMT24 translation tasks~\cite{deutsch2025wmt24expandinglanguagecoverage}. Compared with a chat-interface LLM baseline (GPT-5.3 web), CHORUS reduced completion time by 33.8\%, significantly lowered cognitive workload, and improved translation quality. Participants also found that CHORUS made error inspection clear, which was useful for self-improvement and assessment. This work contributes: (1) formative results on why current LLM translation tools fall short for professionals and insights for improvement; (2) CHORUS, an MQM-inspired multi-agent system that dynamically translates and adapts through multi-modal behavioral signals while scaffolding the translation process and style; and (3) empirical evidence that a multi-agent translation system reduces effort, improves translation quality and speed, and supports more accountable revision. The CHORUS system, including all agent prompts, is publicly available at \url{https://github.com/iamgeorge/CHORUS-OSS}.

%% file: Sections/02_related.tex
\section{Related Work}

\subsection{LLM-Based Translation and Post-Editing}

LLM-based translation work studies prompting, adaptation, and post-editing strategies, including zero-shot prompting, few-shot learning, fine-tuning, prompt design, demonstration quality, and example selection~\cite{raunak2023dissecting,merx2024low,moslem2023adaptive,zhang2023machine,vilar2023prompting}. These methods improve benchmarks and affect document-level behavior, but translation quality remains context-sensitive, and perturbing demonstrations can substantially degrade output~\cite{raunak2023dissecting,zhang2023machine}. Although LLMs can outperform supervised baselines in some directions, commercial accountability remains a concern~\cite{zhu2024multilingual,eschbach2024exploring}, motivating WMT evaluations with professional translators and human evaluation beyond reference-overlap metrics~\cite{freitag2024llms,deutsch2025wmt24expandinglanguagecoverage,freitag2021experts}. LLM-based post-editing can improve MT scores and perceived trustworthiness, and MQM-derived annotations can improve automatic quality metrics such as TER (Translation Edit Rate, number of edits needed to match a reference translation)~\cite{snover2006study}, BLEU (n-gram overlap with reference translations)~\cite{papineni2002bleu}, and COMET (a neural metric trained on human quality judgments)~\cite{rei2020comet,ki2024guiding,lu2025mqm,feng2024improving}. Yet hallucinated edits threaten high-stakes deployment~\cite{raunak2023dissecting, wang2025hidden}, and it remains unclear how annotations should guide concrete revision decisions~\cite{ki2024guiding}. This motivates human-in-the-loop workflows where translators retain oversight over quality and accountability~\cite{freitag2021experts,wang2022non}. Iterative and multi-agent LLM work also shows both promise and limits: self-refinement and iterative translation refinement can improve outputs, fluency, and naturalness, but can complicate metric interpretation and amplify model self-bias unless external feedback is introduced~\cite{madaan2023self,chen2024iterative,xu2024pride}. Multi-agent systems use specialized roles for translation production or evaluation~\cite{wu2024transagents,briva2025ai,zhang2025himate,zhang2025diting}, and interactive MT emphasizes translator control~\cite{foster1997target,langlais2000transtype}; however, prior systems often focus on agent specialization, agent-to-agent refinement, or traditional interactive MT rather than close human coordination during multi-dimensional revision~\cite{alabau2013casmacat,huang2021transmart,koponen2016machine}. CHORUS builds on this space by combining LLM revision, specialized agent support, and translator oversight.

\subsection{Quality Evaluation in Translation}

Translation quality is assessed through automatic metrics and human evaluation, but evaluation can be misleading without explicit error analysis, especially for strong MT systems where differences are subtle~\cite{marie2021scientific,graham2019translationese,zhao2020limitations,laubli2020set,song2025enhancing}. Prior work therefore argues for structured error types and severities~\cite{freitag2021experts,lommel2018metrics}. Automatic metrics include BLEU, which depends on reference similarity~\cite{papineni2002bleu,reiter2018structured,callison2006re}; TER, which estimates revision effort~\cite{snover2006study,snover2009ter,przybocki2006edit}; and COMET, trained on human judgments such as Direct Assessment, HTER/TER-style judgments, and MQM annotations~\cite{rei2020comet,bentivogli2018machine,snover2006study,lommel2014multidimensional}. Human evaluation protocols and quality frameworks include Direct Assessment, continuous scoring, the Dynamic Quality Framework, SAE J2450, the LISA QA Model, and Multidimensional Quality Metrics (MQM)~\cite{bentivogli2018machine,graham2013continuous,chatzikoumi2020evaluate,gorog2014quality,schutz1999deploying,rws_lisa_qa_model,lommel2014multidimensional}. Following prior work, we use MQM because it provides a fine-grained taxonomy of translation errors and is widely used in professional, research, operational quality-control, and agent-based evaluation settings~\cite{lommel2014using,freitag2021results,freitag2021experts,lu2025mqm,lommel2024multi}. In CHORUS, MQM defines the quality concerns available in the interface and the seven dimensions summarized in Table~\ref{tab:mqm-dimensions}.

\subsection{Adaptive Translation Systems}

Adaptive translation systems treat translation as an interactive process: the system proposes translations, humans correct or rate them, and the system updates to reduce future effort~\cite{freitag2021experts,ortiz2011interactive,peris2019online,wang2024synslator}. Mixed-initiative and interactive MT have long offered alternatives to pure post-editing through target-text-mediated interaction, completions compatible with translator input, links between human effort and machine learnability, and feedback datastores for later translations~\cite{green2015natural,foster1997target,langlais2000transtype,green2014human,wang2022non}. Recent adaptive work collects fine-grained edits, prefix constraints, accept/reject actions, segment-level post-edits, and structured error tags to reduce editing cost and make feedback reusable~\cite{yamaguchi2024automatic,dong2019editnts,knowles2016neural,lam2019interactive,ki2024guiding,yuksel2025efficient,gois2019translator2vec,wang2025maats}. LLM-enabled systems can produce useful edits but still require validation and human oversight~\cite{raunak2023leveraging}; iterative self-refinement can improve fluency and naturalness while complicating metric interpretation~\cite{madaan2023self,chen2024iterative,pillutla2021mauve}. This points to a design gap for human-centered multi-agent translation: adaptation should respond not only to user feedback, but also to the effort and commitment behind that feedback.

\begin{table}[t]
    \centering
    \small
    \caption{Participant profile for the formative study.}
    \Description{A five-column table listing six formative-study participants, their organizational side, domain, years of experience, and gender.}
    \label{tab:formative-participants}
    \begin{tabular}{@{}lllll@{}}
        \toprule
        Expert & Side & Domain & Experience & Gender \\
        \midrule
        E1 & Client & Game Localization & 12+ & Female \\
        E2 & Client & Marketing & 6+ & Male \\
        E3 & Client & Government & 8+ & Female \\
        E4 & Vendor & Medical & 10+ & Female \\
        E5 & Vendor & Game Localization & 7+ & Male \\
        E6 & Vendor & Chip Design & 11+ & Female \\
        \bottomrule
    \end{tabular}
\end{table}

%% file: Sections/03_formative_study.tex
\section{Formative Study}

\subsection{Participants and Data Collection}
We interviewed six professional translators (E1--E6), evenly split between client-side and vendor-side roles. Each participant had at least 6 years of professional practice across diverse translation domains, including game localization, marketing, government, medical translation, and chip design (Table~\ref{tab:formative-participants}). All interviews were conducted remotely via Zoom, and sessions were recorded and transcribed. Interviews averaged 42 minutes. The study was reviewed by our institution's IRB and granted Exempt status, and all participants provided informed consent prior to the interview.

\subsection{Procedure}
We conducted semi-structured interviews consisting of three parts:
\textbf{Background and tool usage:} Participants were asked about the AI tools they use and their satisfaction with these tools (including reasons). \textbf{Think-aloud translation task:} Participants were given a sentence to translate in real time while verbalizing their thought process and explaining their edits. \textbf{Open-ended design reflection:} Participants were asked to imagine and describe their ideal translation tool while thinking aloud. The full interview protocol is provided in Appendix.

We analyzed the interview transcripts and notes using thematic analysis~\cite{braun2006using}. Two authors conducted open coding to identify recurring issues and breakdowns in current translation workflows. The authors then compared and discussed their codes, resolved disagreements through discussion, and grouped related codes into higher-level themes. This analysis led to three recurring challenges that informed the design of CHORUS.

\subsection{Challenges in Current AI Translation Tools}
\textbf{C1. Single-LLM rewrites obscure distinct translation quality dimensions.}
Participants described professional revision as a process of balancing multiple quality dimensions within the same sentence, including accuracy, terminology, fluency, and style~(E1, E2, E5). However, current AI tools often return a single broad rewrite, making it difficult for translators to see which quality dimension the system addressed and whether the revision introduced trade-offs elsewhere. As one participant explained, ``It often rewrites everything at once. I cannot tell if it is fixing terminology or just changing the style, so I end up going back and rechecking''~(E2). Across interviews, all six experts reported general-purpose AI chatbots (GPT web) as their primary day-to-day AI aid, while CAT tools were reserved for projects with established translation memories. 

\textbf{C2. Current AI tools do not carry forward translators' quality adjustments.} 
Participants emphasized that translation decisions unfold through revision: a sentence may evolve through several versions, some segments may be repeatedly reconsidered, and translators may concentrate more attention on one quality concern than another~(E1, E2). These process traces matter because not all edits carry the same weight. A terminology correction in medical content may reflect a domain constraint, while a fluency revision may matter more in consumer-facing text~(E2, E6). However, current AI tools rarely preserve these signals across the workflow. Translators must repeatedly tell the system what to prioritize, as in E3's example: ``This part has been revised three times. Don't change it, but you can still improve the surrounding sentences.''

\textbf{C3. Current AI tools provide little support for reviewing and justifying translation decisions.}
Finally, participants wanted more structured guidance for checking what had been addressed during revision~(E1, E5). They described needs such as reviewing style guides~(E1), identifying critical issues~(E1, E3), checking domain-specific constraints~(E4), and understanding why a suggestion was made~(E5). These needs were tied to professional accountability: translators often have to justify and defend decisions to supervisors, clients, or other stakeholders~(E3, E4, E5). Current AI tools offer suggestions, but provide limited support for seeing which quality dimensions have been checked, which concerns may still need attention, and how a final decision can be explained.

\subsection{Design Rationale}
\textbf{D1. Operationalize MQM as separable agent roles.}
Because LLM-based translation support is context-sensitive, a single-LLM interface can merge several quality goals into one opaque rewrite and become difficult for translators to inspect. All participants acknowledged MQM as an authoritative framework for multi-dimensional thinking. We therefore decompose revision support into MQM-aligned agents, each focused on one quality lens, so translators can inspect dimension-specific suggestions and coordinate trade-offs before deciding on the final translation.

\textbf{D2. Capture and use editing history to reduce repeated MQM-specific quality adjustments.}
Professional translators often need to steer the balance among quality dimensions across many segments, such as preserving terminology, adjusting fluency, or maintaining a client-specific style. In current LLM workflows, these preferences often have to be restated through repeated prompts, which adds effort to the revision process. We should use the translator's interaction history to ease this work. We need to introduce a novel mechanism with confirmed edits and revision traces to help the system recognize which quality adjustments the translator has already made and carry them forward into later suggestions.

\textbf{D3. Support structured review through revision history and quality coverage.}

    We should help translators review their own revision process without turning reflection into a separate task. The interface should surface useful traces of the work already done: recurring edits and preferences from revision history, how attention has been distributed across MQM dimensions, and which quality concerns may still deserve review. This gives translators material for checking coverage and explaining decisions while keeping final judgment with the user.

\begin{table*}[t]
    \caption{Seven MQM-aligned dimensions operationalized in CHORUS, adapted from MQM error typologies~\cite{lommel2018metrics}.}
    \label{tab:mqm-dimensions}
    \Description{A three-column table summarizing seven MQM-aligned dimensions used in CHORUS: Accuracy, Terminology, Fluency, Style, Audience Appropriateness, Locale Convention, and Design and Markup, with each dimension's primary focus and the type of issue its expert checks.}
    \small
    \centering
    \renewcommand{\arraystretch}{1.15}
    \begin{tabular}{@{}>{\raggedright\arraybackslash}p{0.19\textwidth} >{\raggedright\arraybackslash}p{0.19\textwidth} >{\raggedright\arraybackslash}p{0.54\textwidth}@{}}
        \toprule
        \textbf{Dimension} & \textbf{Primary focus} & \textbf{What the expert checks} \\
        \midrule
        Accuracy & Meaning preservation & If the translation conveys source meaning without omissions, additions, or distortions. \\
        Terminology & Lexical consistency & If domain-specific terms and glossary entries are used correctly and consistently. \\
        Fluency & Linguistic well-formedness & If the sentence is grammatical, natural, and readable in the target language. \\
        Style & Tone and register & If the wording matches the intended voice, formality, and conventions of the task. \\
        Audience Appropriateness & Reader fit & If the translation is suitable for the intended audience and expertise level. \\
        Locale Convention & Regional adaptation & If the translation follows culturally specific conventions, formats, and templates. \\
        Design and Markup & Structural integrity & If formatting, line breaks, placeholders, and layout-sensitive elements are correct. \\
        \bottomrule
    \end{tabular}
\end{table*}

%% file: Sections/04_system.tex
\section{CHORUS System}

\begin{figure*}[t]
    \centering
    \includegraphics[width=0.9\textwidth]{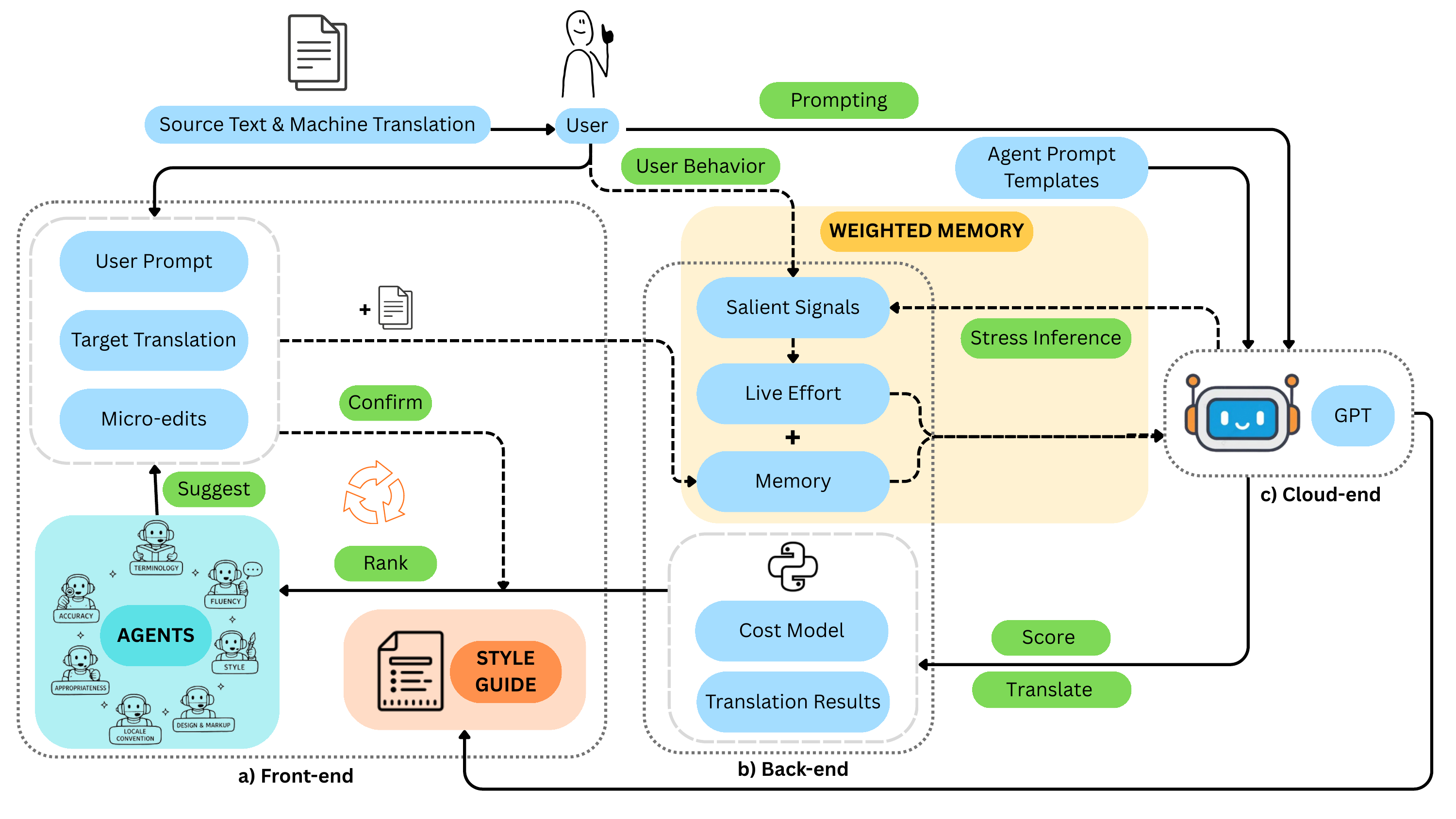}
    \caption{System flow chart of CHORUS workflow integrating context, draft, goals, agents, revisions, effort, and memory.}
    \Description{Flow diagram of CHORUS in three parts: a front-end where the user works with the source text, target translation, micro-edits, seven agents, and a style guide; a back-end where a weighted memory combines salient signals, live effort, and memory alongside a cost model and translation results; and a cloud-end GPT model. Arrows show user behavior and confirmed edits feeding the weighted memory, which together with prompts drives the GPT model to score, rank agents, and generate suggestions that update the translation and style guide.}
    \label{fig:system-flow}
\end{figure*}

\begin{figure*}[t]
    \centering
    \includegraphics[width=0.9\textwidth]{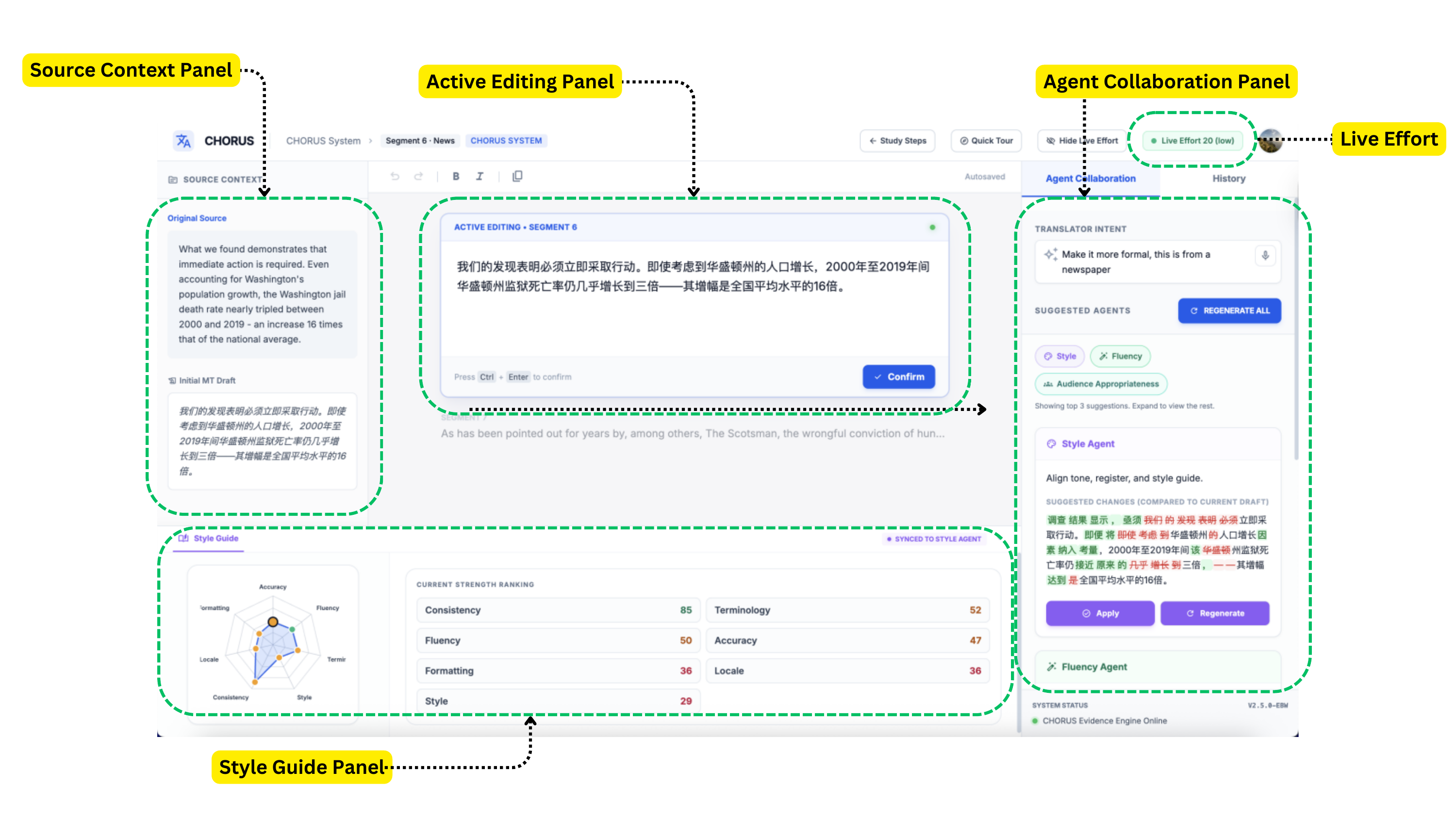}
    \caption{Overview of CHORUS. Source Context Panel (A) shows the source and initial translation. The Active Editing panel (B) is the main workspace for revision. The Agent Collaboration panel (C) provides suggestions from MQM agents. The Live Effort (D) indicates real-time translator effort. The Style Guide panel (E) provides learned preferences and user feedback.}
    \Description{The CHORUS interface includes five panels: source context with original and translated text, an editing workspace, agent-based suggestion controls, a live effort indicator based on interaction traces, and a style guide summarizing learned preferences.}
    \label{fig:system-overview}
\end{figure*}

\subsection{Human-AI Multi-Agent Collaboration}

\subsubsection{Synchronization among multiple agents}
Separating quality dimensions introduces a coordination problem: suggestions from multiple agents must remain anchored to the same current draft. CHORUS therefore synchronizes agent outputs after each user edit. Each agent's response is represented as a token-level difference against the latest draft using the Longest Common Subsequence algorithm~\cite{kruskal1983overview, wagner1974string, hunt1977fast}. CHORUS applies the minimal patch and immediately re-renders agent suggestions (see Fig.~\ref{fig:system-flow}), keeping dimension-specific feedback aligned with the translator's current text.
            
\subsubsection{Ranking agents}
 
Since several dimensions may be relevant at once, CHORUS ranks agents to reduce attentional load while preserving access to the full MQM space. The system obtains an initial relevance score for each agent from the translator's high-level goal and current context. Subsequent interactions update each agent's score, so agents that better match the translator's current focus become more visible over time. This serves as an attention-management mechanism: the most relevant agents are ranked according to the current context while keeping other MQM dimensions always available.

\subsubsection{Error handling}
Hallucinations are known to cause errors in LLMs' responses~\cite{raunak2023dissecting,wang2025hidden}. CHORUS employs a list of ``bad examples'' whenever the user identifies an error in LLM output. During regeneration, CHORUS commands the LLM to avoid similar mistakes cached in the bad example list. Additionally, users can regenerate any single agent's output if unsatisfactory.

\subsection{Effort-Aware Memory}
\label{sec:live-effort}
To help the system personalize toward users' habits, we introduce an effort-aware memory mechanism that converts users' editing behavior into weighted memory for prompt adaptation. Live Effort estimates which edits required more translator effort, Micro-Edits preserve what was changed and where the change occurred, and Memory uses these effort-weighted edit records to generate prompt guidance for the AI agents. Together, these components allow CHORUS to remember which edits matter the most and adjust future prompts accordingly.

\textbf{Live Effort} Following Krings and Stasimioti~\cite{krings2001repairing,stasimioti2020translation}, CHORUS models live effort at three levels: temporal, technical, and cognitive. \textit{Temporal effort} is defined by two metrics: an initial pause that reflects the time spent reading and understanding the source text and a total edit duration that captures the time to refine. \textit{Technical effort} is measured through keystroke counts (deletion and cursor movement). \textit{Cognitive effort} captures the deeper reasoning processes, such as the number of redos and stress levels, which are inferred by ChatGPT-5.3 using appraisal theories of emotion (difficulty, ambiguity, risk, and controllability)~\cite{scherer2014nature, perrez1996computer}. To combine these three metrics into one live effort score, we use a linear effort model~\cite{turchi2013coping,alabbas2025weighted,stasimioti2020translation} and join these metrics into the final score.

\textbf{Micro-Edits}
capture deletions or replacements during editing. This design allows the system to preserve not only \textit{what} the translator changed, but also \textit{where} the change occurred. It can also be used to infer \textit{why} the change was made and, when combined with the system's memory component, \textit{how much effort} it required.

\textbf{Memory}
CHORUS uses a weighted memory to reduce the need for repeated edits. Inspired by importance-aware retrieval~\cite{zhong2024memorybank,park2023generative}, CHORUS uses seven memory buckets to match up with seven AI agents' editing history. The memory contains a snapshot of micro-edits, target translation, and user prompt whenever a change is made. Non-agent edits such as modifying the output sentence directly are stored in a general memory bucket. To personalize CHORUS, we use the stored memory to adjust the prompt template to fit the professional translator's idiosyncratic traits. This begins with ranking micro-edits using the live effort algorithm. After ranking, CHORUS injects the top-five micro-edits as few-shot examples into a prompt template to generate a weighted memory, personalizing what matters most to users without restating their needs. Appendix shows a complete example and prompt templates.

\subsection{Translational Scaffolding}

We instrument \textit{Live Style Guide} to show translational scaffolding by surfacing the learned preferences, recurring correction patterns, and dimension-level strengths as a spider graph using the weighted memory. This helps users to see which of the MQM dimensions are more frequently used and their personalized feedback on their performance. This feedback is created based on the LLM's response using the MQM website's suggestions~\cite{mqm_website}. This way, the system helps professional translators to recognize their blind spots over time and reflect on their strengths. 

%% file: Sections/05_user_study.tex
\section{Evaluation}

\subsection{Experiment Design}
The evaluation used English-to-Chinese sentence-level data from WMT24~\cite{deutsch2025wmt24expandinglanguagecoverage}. We sampled 10 sentences from each of four domains: \textit{literary}, \textit{news}, \textit{social}, and \textit{speech}. WMT24 reference translations were retained for quality evaluation. In the \textbf{Baseline} condition, participants revised translations while using the GPT-5.3 web~\cite{openai2026gpt53} as AI support. Chat history was cleared between trials. In the \textbf{CHORUS} condition, they worked with the CHORUS interface (see Figure~\ref{fig:system-overview}), which contains the source text, an editable machine-translated draft, and access to seven AI agents. In both conditions, initial drafts were produced by GPT-5.3. Condition order was counterbalanced using a pre-generated table, and to avoid learning effects, no identical sentence appeared across conditions or trials. 

CAT tools were not used because they rely on accumulated translation memories and termbases unavailable for our tasks. Other systems were also unsuitable as end-to-end baselines: CASMACAT~\cite{alabau2013casmacat} predates LLM-based assistance and was not evaluated on English--Chinese, TranSmart~\cite{huang2021transmart} is a closed commercial service, and MQM-APE and HiMATE~\cite{lu2025mqm,zhang2025himate} provide evaluation methods rather than interactive translation systems.

\subsection{Participants}

We recruited 30 licensed professional English--Chinese translators who provided proof of certification. Participants were 21 to 50 years old ($M{=}28.9$, $SD{=}6.1$) and reported 3 to 21 years of translation experience ($M{=}5.4$, $SD{=}4.9$). They reported frequent use of AI translation tools (median $= 4/5$) and moderate-to-high familiarity with CAT tools (median $= 3.5/5$). Each participant was assigned two out of the four possible WMT24 domains and paired with conditions. For each domain, participants edited 10 sentences (10 trials), and each domain contained the two conditions. As a result, each participant conducted 10 sentences per condition, 20 per domain, and 40 in total. The overall frequency of domains was balanced across all participants. For example, we assigned $P_n$ domains A and B and $P_{n+1}$ domains C and D. This pattern was alternated.

\subsection{Procedure and Measures}

After participants signed the consent form, the experimenter explained the tasks and walked them through the interface and how to perform the trial. Participants had five minutes to practice. Once ready, they were assigned one of the conditions and switched to the other condition upon completing the first. The NASA-TLX survey~\cite{hart1988development} was used to assess cognitive load across conditions. At the end of the experiment, a semi-structured interview followed to collect qualitative feedback and their impressions of the two conditions. Sessions lasted approximately 100 minutes, and participants were compensated ¥100. 

We collected interaction logs, outcome translations, and self-report data. Timed performance was computed from active editing duration, excluding idle periods. Perceived workload was measured with NASA-TLX on the original 0--100 scale, and open-ended responses were collected. To assess quality, we used COMET~\cite{rei2020comet} and BLEU~\cite{papineni2002bleu} automatic evaluation metrics as a means of automatic comparison as they come with ``ground-truth'' labeling. Further, we asked three professional translators with more than 6 years of experience to compare the translated sentence with the original. 

%% file: Sections/06_results.tex
\section{Results}

\subsection{Efficiency, Effort, and Workload}

\begin{figure}[t]
    \centering
    \includegraphics[width=\columnwidth]{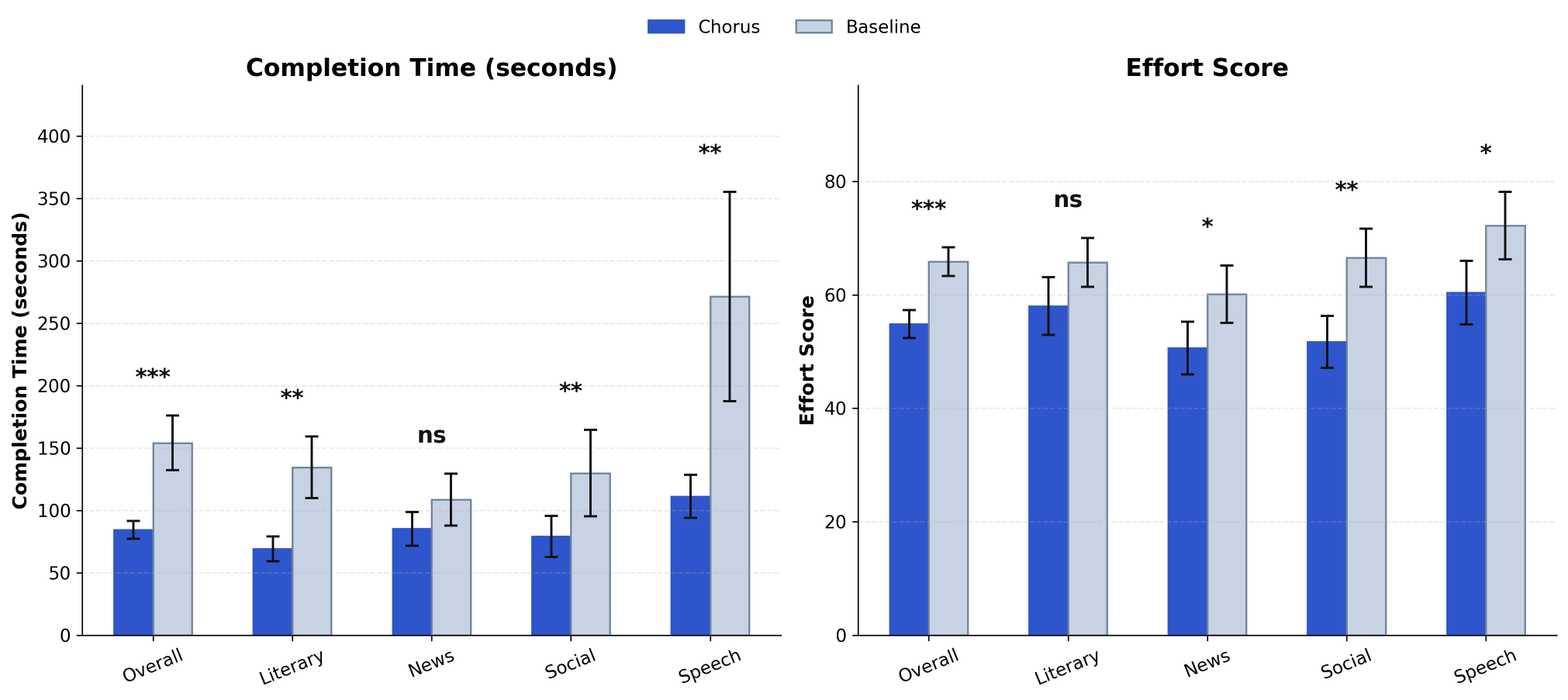}
    \caption{Completion time and live effort scores for CHORUS and Baseline across domains.}
    \Description{Two grouped bar charts comparing CHORUS and Baseline on completion time and Live Effort score, overall and for the literary, news, social, and speech domains. CHORUS bars are lower in every comparison, with error bars and significance markers.}
    \label{fig:completion-time-effort}
\end{figure}

A paired $t$-test on log-transformed completion time showed that CHORUS significantly reduced completion time compared with Baseline ($t(59) = -5.35$, $p < .001$). On the original scale, the geometric-mean time ratio of CHORUS over Baseline was 0.662, 95\% CI [0.567, 0.772], indicating 33.8\% faster task completion on average. Domain-level comparisons followed the same direction, with the largest reductions in literary and speech tasks (Fig.~\ref{fig:completion-time-effort}).

CHORUS also produced significantly lower Live Effort scores than Baseline ($t(59) = -5.81$, $p < .001$). The average effort score was 54.85 ($SD = 19.14$) for CHORUS and 65.84 ($SD = 19.61$) for Baseline, a reduction of about 11 points. Mixed-effects analyses found no significant omnibus domain effect for completion time, $\chi^2(3) = 5.45$, $p = .142$, or effort, $\chi^2(3) = 6.00$, $p = .112$, suggesting that the CHORUS advantage was not driven by a single domain.

\begin{figure}[t]
    \centering
    \includegraphics[width=\columnwidth]{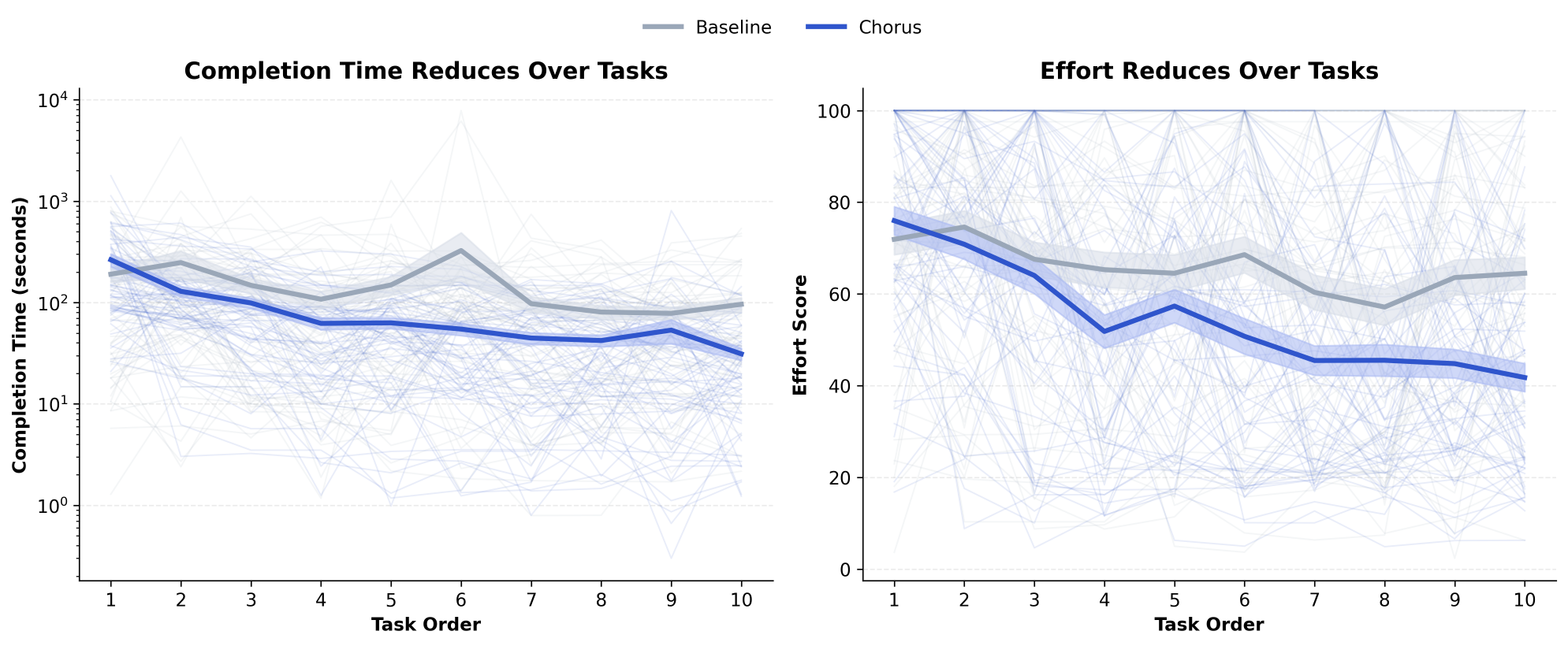}
    \caption{Task-order trajectories for completion time and live effort for two conditions. Completion time is log-scaled.}
    \Description{Two line charts showing completion time on a log scale and Live Effort score across task order one to ten. Faint lines show individual participants and bold lines show condition means; the CHORUS line starts near Baseline, declines more steeply, and remains below Baseline from the second task onward.}
    \label{fig:loop-reduction-overview}
\end{figure}

\begin{figure}[t]
    \centering
    \includegraphics[width=\columnwidth]{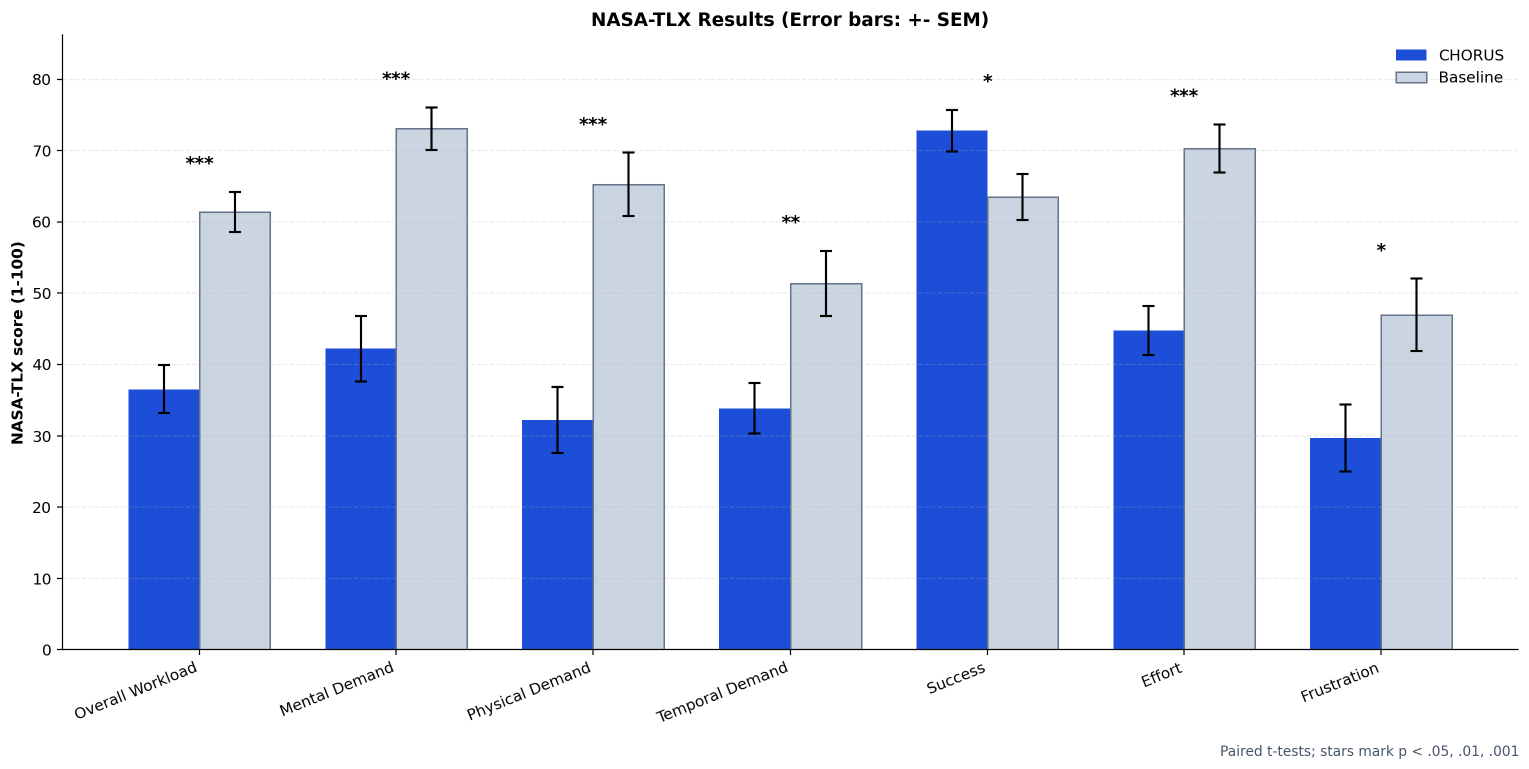}
    \caption{NASA-TLX scores for CHORUS and Baseline.}
    \Description{Grouped bar chart of NASA-TLX scores for CHORUS and Baseline across overall workload, mental demand, physical demand, temporal demand, success, effort, and frustration. CHORUS scores lower on all workload sub-scales and higher on perceived success, with significance markers.}
    \label{fig:nasa-tlx}
\end{figure}

Task-order analyses further suggested that participants settled into more efficient interaction patterns with CHORUS over time (Fig.~\ref{fig:nasa-tlx}). For completion time, the interaction between task order and CHORUS was significant, $b = -0.146$, $z = -6.66$, $p < .001$, with a significant mean slope difference against Baseline, $-0.147$, $t(59) = -6.12$, $p < .001$. Live Effort showed the same pattern: the CHORUS interaction was significant, $b = -2.444$, $z = -5.67$, $p < .001$, with a significant mean slope difference of $-2.448$, $t(59) = -5.14$, $p < .001$ (Fig.~\ref{fig:loop-reduction-overview}).

\subsection{Translation Quality and Adaptivity}

CHORUS improved final translation quality relative to Baseline. Measured against human-translated WMT references, mean BLEU increased from 34.90 under Baseline to 37.98 under CHORUS, a mean paired difference of 3.08 points ($t(29) = 3.16$, $p = .0036$; Wilcoxon $p = .0081$; Hedges' $g = 0.56$). Mean COMET increased from 0.837 to 0.852, a mean paired difference of 0.015 ($t(29) = 3.51$, $p = .0015$; Wilcoxon $p = .0019$; Hedges' $g = 0.62$). For both metrics, 73.3\% of participants had higher scores under CHORUS than Baseline.

\begin{figure}[t]
    \centering
    \includegraphics[width=\columnwidth]{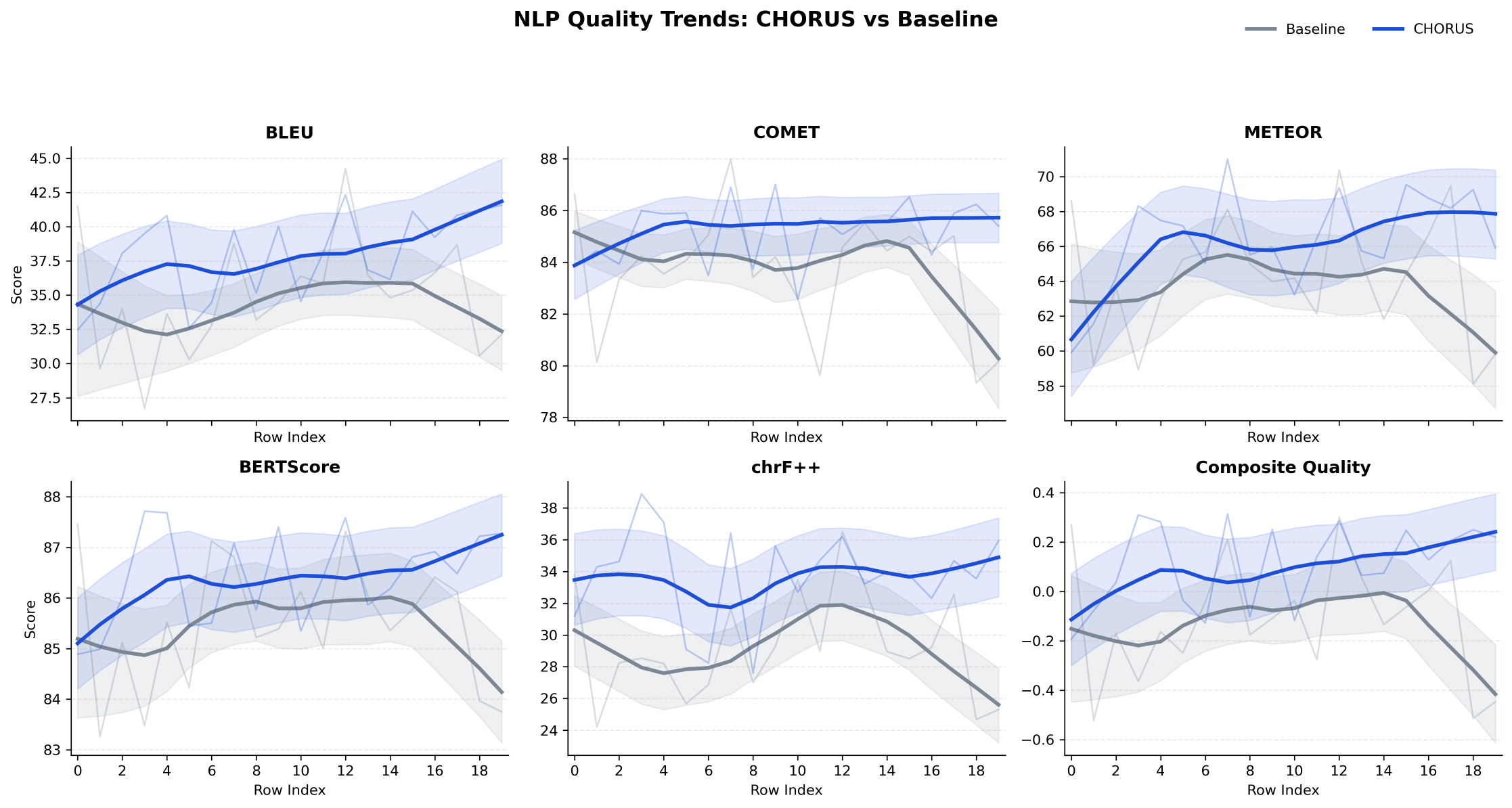}
    \caption{Task-order trends in translation-quality metrics under CHORUS and Baseline.}
    \Description{Six line charts showing BLEU, COMET, METEOR, BERTScore, chrF++, and a composite quality index over task order, with smoothed trend lines and confidence bands. CHORUS trends upward or stays stable while Baseline remains flat or declines in later tasks.}
    \label{fig:adaptivity-trend}
\end{figure}

Quality trends over task order provided additional evidence of adaptation (Fig.~\ref{fig:adaptivity-trend}). 
BLEU increased over rows under CHORUS (slope $= 0.297$, $p = .014$), as did METEOR~\cite{banerjee2005meteor} (slope $= 0.00236$, $p = .024$), while Baseline remained largely flat on these metrics. A composite quality index was flat for Baseline (slope $= 0.00062$, $p = .923$) and showed an upward trend for CHORUS (slope $= 0.0119$, $p = .059$). BLEURT~\cite{sellam2020bleurt} also suggested that Baseline quality declined over rows (slope $= -0.00830$, $p < .001$), while CHORUS showed a smaller, non-significant decrease (slope $= -0.00295$, $p = .101$).

\begin{figure}[t]
    \centering
    \includegraphics[width=\columnwidth]{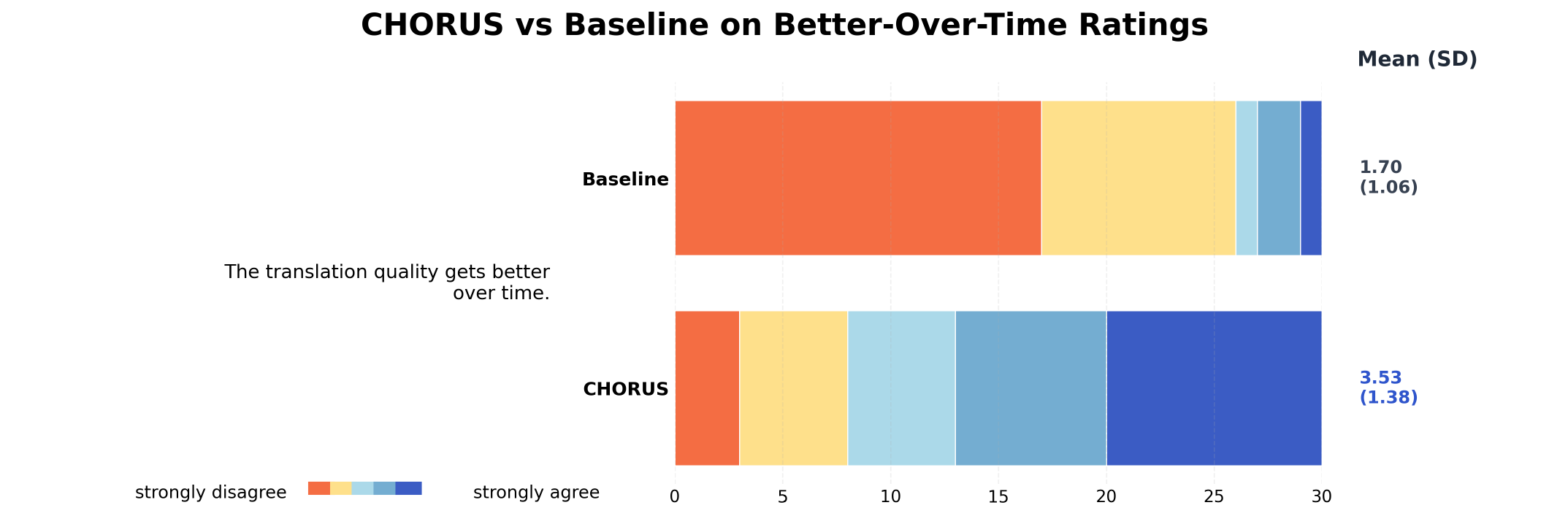}
    \caption{Human expert ratings of whether translation quality improved over time under CHORUS and Baseline.}
    \Description{Horizontal stacked bar chart of expert agreement with the statement that translation quality gets better over time. Baseline ratings concentrate toward strong disagreement with a mean of 1.70, while CHORUS ratings concentrate toward agreement with a mean of 3.53.}
    \label{fig:human-eval}
\end{figure}

Human expert ratings showed the same direction. Three translation experts rated whether translation quality improved over time. CHORUS received a mean rating of 3.53 ($SD = 1.38$), whereas Baseline received 1.70 ($SD = 1.06$), with Baseline ratings concentrated toward disagreement and CHORUS ratings showing more agreement (Fig.~\ref{fig:human-eval}).

\subsection{Structured Revision and Reflection}

\begin{figure}[t]
    \centering
    \includegraphics[width=\columnwidth]{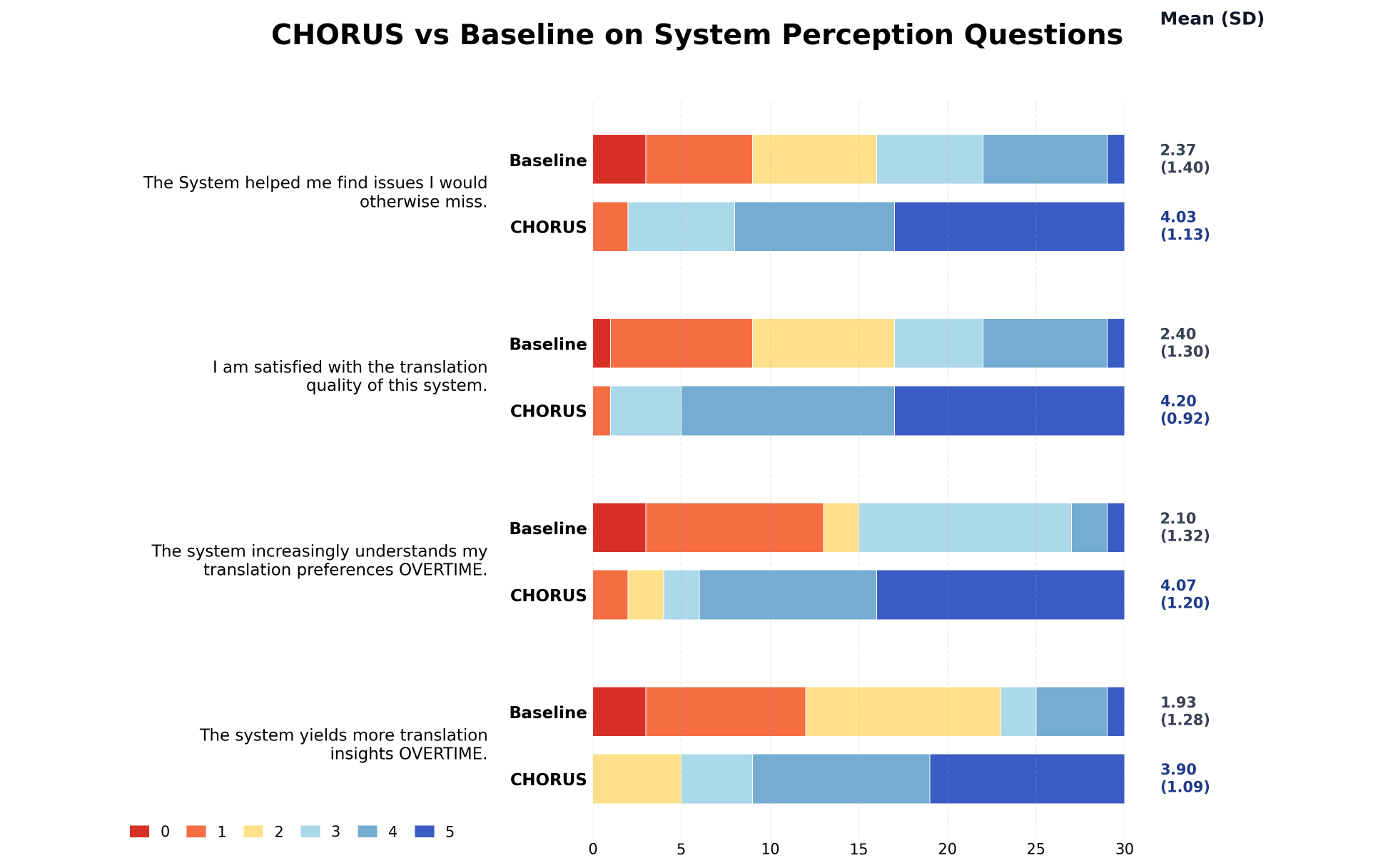}
    \caption{Participant responses to perception questions.}
    \Description{Horizontal stacked bar charts of participant ratings for four perception statements, each comparing CHORUS and Baseline with means and standard deviations. CHORUS receives higher agreement on every statement.}
    \label{fig:perception-questions}
\end{figure}

Open-ended responses helped explain why CHORUS reduced effort and improved quality. Participants attributed the lower effort to targeted support, clearer problem visibility, and less need to manually formulate prompts. They described CHORUS as making issue-specific feedback easier to inspect than broad chatbot responses, and several participants noted that focusing on one quality dimension at a time reduced the burden of mentally tracking accuracy, terminology, fluency, and style together (Fig.~\ref{fig:perception-questions}).

Participants also described CHORUS as supporting more deliberate quality decisions. They reported that dimension-specific suggestions made trade-offs more explicit, helped them catch issues they might otherwise miss, and supported verification before confirming a sentence. In contrast to a static chatbot, participants described CHORUS as increasingly reflecting their ongoing edits, preferred correction patterns, and habitual areas of focus, reducing the need to repeatedly restate the same priorities.

\begin{figure}[t]
    \centering
    \includegraphics[width=\columnwidth]{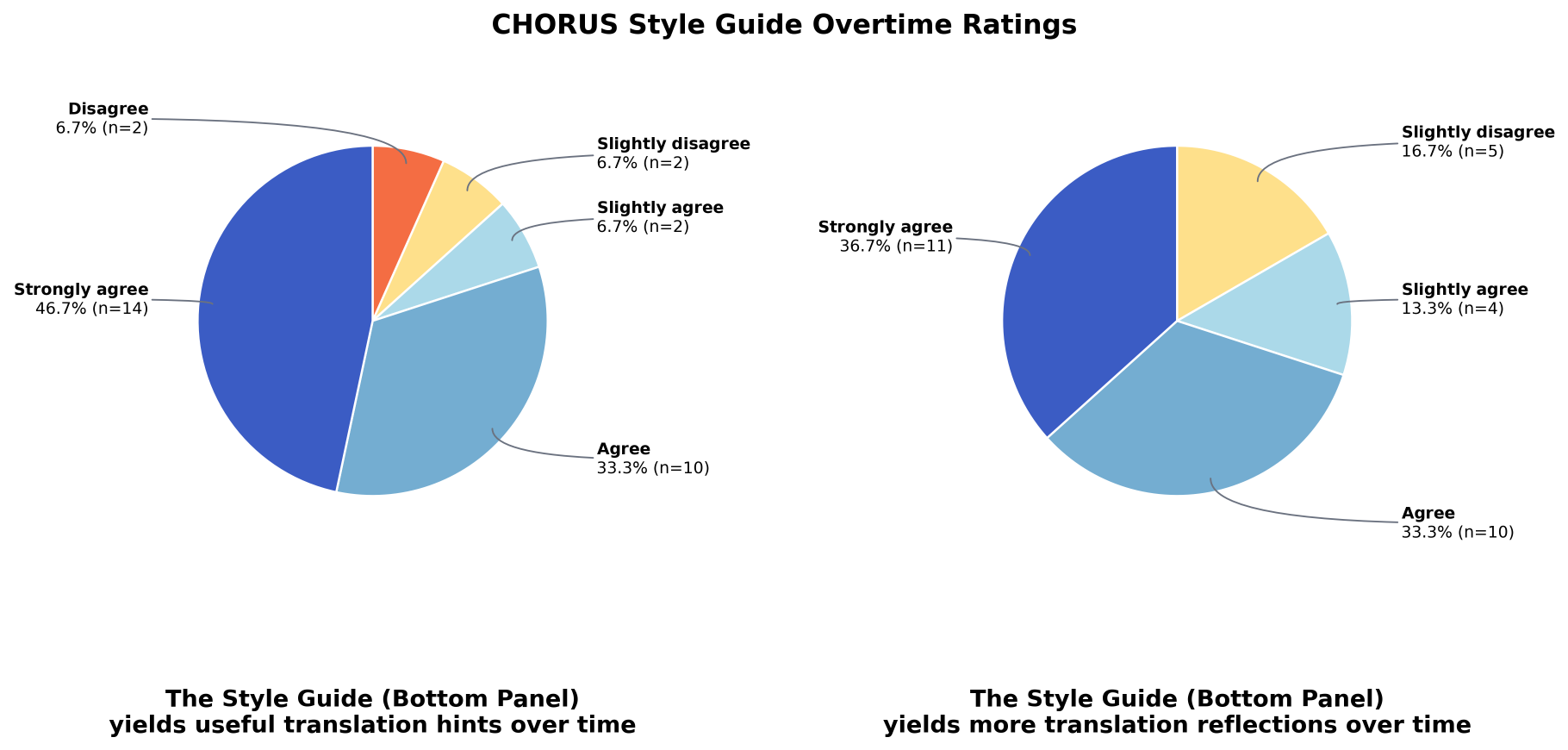}
    \caption{Style-guide dimension distributions over time.}
    \Description{Two pie charts of participant agreement that the Style Guide yields useful translation hints over time and more translation reflections over time. In both charts, most participants agree or strongly agree.}
    \label{fig:style-guide-overtime}
\end{figure}

The Style Guide and radar chart supported structured review by making revision history and MQM coverage visible. Participants used the radar chart to see which dimensions received more attention and which were underrepresented, and many reported using the visualization to rebalance their focus in later revisions. The Style Guide also helped participants understand MQM dimensions, identify possible misclassifications in their own edits, and justify decisions with more confidence (Fig.~\ref{fig:style-guide-overtime}).

%% file: Sections/07_discussion.tex
\section{Discussion}

The results suggest that CHORUS improves professional translation not by replacing translator judgment, but by reorganizing LLM assistance around how translators revise. MQM agents made quality concerns more visible and easier to inspect, reducing the need to repeatedly prompt, compare broad responses, and mentally track several revision goals at once. This shift may help explain the 33.8\% reduction in completion time, the lower Live Effort, and the lower NASA-TLX workload (RQ1). Because CHORUS differs from the baseline in both its multi-agent backend and its purpose-built interface, we interpret these gains as the effect of the integrated workflow rather than of the agent architecture alone. It also aligns with prior work on mixed-initiative translation, where useful automation supports professional control rather than removing it~\cite{foster1997target,green2015natural,huang2021transmart}.

CHORUS improved translation metrics (BLEU $+3.08$, COMET $+0.015$) and expert ratings (RQ2). One explanation is that participants could evaluate whether a suggestion improved accuracy, fluency, or another MQM dimension before committing to it. Making trade-offs inspectable reduced the risk of accepting suboptimal rewrites and supported more deliberate final decisions.

CHORUS also supported reflection beyond sentence-level editing (RQ3), complementing the inspectability and reduced re-prompting that participants highlighted in their qualitative feedback. The Style Guide and radar chart made revision history and quality coverage visible, helping participants notice where their attention was concentrated, identify possible blind spots, and explain decisions using shared quality categories. 

More broadly, CHORUS illustrates a division of labor for human-centered multi-agent systems. Agents can surface dimension-specific issues, organize alternatives, and accumulate interaction history, while the professionals remain responsible for interpreting context, weighing trade-offs, and making the final decision. This pattern may generalize to other revision-heavy domains where people coordinate multiple specialized AI agents under shifting priorities. 

%% file: Sections/08_limitation_n_future.tex
\section{Limitations and Future Work}

Our evaluation assessed CHORUS as an integrated workflow because the goal of evaluation was to holistically understand the efficacy of the system for professional translation. Efficacy of individual components remains the future work.

In addition, our evaluation was limited to sentence-level English--Chinese tasks from WMT24. Other language pairs, and domain-specific effort signals remain to be tested, and we hope to expand the user study through multilingual and document-level studies in future work.

%% file: Sections/09_conclusion.tex
\section{Conclusion}
This paper presented CHORUS, an MQM-aligned multi-agent workspace for professional translators. CHORUS separates revision into seven inspectable dimensions and integrates users' interaction history to reduce repeated MQM-specific adjustments, and visualizes revision patterns through an MQM review interface. In a study with 30 professional translators, CHORUS reduced completion time, effort, and workload while improving translation quality evaluated via both expert ratings and existing methods. Findings suggest that CHORUS can better support professional work by preserving translators' control and revealing quality concerns for inspections.

%% file: Sections/10_acknowledgement.tex
\section{Safe and Responsible Innovation Statement}
CHORUS explores a human-accountable paradigm for multi-agent interaction in professional translation, where AI support remains explainable, inspectable, and subject to human judgment. Responsible deployment should protect confidential materials, audit uneven performance across users and domains, and prevent over-reliance on agent suggestions in high-stakes settings.